\documentclass[11pt,a4paper]{article}
\pdfoutput=1
\usepackage{jinstpub}
\usepackage{subfig}
\usepackage{lineno}
\usepackage{booktabs}

\title{High-resolution tracking in a GEM-Emulsion detector}
\author[a]{A.~Alexandrov}
\author[c]{, G.~Bencivenni} 
\author[c]{, M.~Bertani} 
\author[a,b,1]{, A.~Buonaura \note{Corresponding author.}}
\author[c]{, C.~Capoccia} 
\author[d]{, G.~Cibinetto}
\author[a,b]{, G.~De Lellis}
\author[c]{, E.~De Lucia}
\author[a,b]{, A.~Di Crescenzo}
\author[c]{, D.~Domenici}
\author[d]{, R.~Farinelli}
\author[c]{, G.~Felici}
\author[e]{, N.~Kitagawa}
\author[e]{, M.~Komatsu}
\author[c]{, G.~Morello}
\author[e]{, K.~Morishima}
\author[c]{, M.~Poli Lener}
\author[a]{, V.~Tioukov}
	\affiliation[a]{INFN Sezione di Napoli, \\Napoli, Italy}
	\affiliation[b]{Dipartimento di Fisica dell'Universit\`a Federico II di Napoli, \\Napoli, Italy}
	\affiliation[c]{Laboratori Nazionali dell'INFN di Frascati, \\Frascati, Italy}
	\affiliation[d]{INFN Sezione di Ferrara, \\Ferrara, Italy}
	\affiliation[e]{Nagoya University, \\Nagoya, Japan}
	\emailAdd{annarita.buonaura@cern.ch}

\abstract{SHiP (Search for Hidden Particles) is a beam dump experiment proposed at the CERN SPS aiming at the observation of long lived particles very weakly coupled with ordinary matter mostly produced in the decay of charmed hadrons. The beam dump facility of SHiP is also a copious factory of neutrinos of all three kinds and therefore a dedicated neutrino detector is foreseen in the SHiP apparatus. The neutrino detector exploits the Emulsion Cloud Chamber technique with a modular structure, alternating walls of target units and planes of electronic detectors providing the time stamp to the event. GEM detectors are one of the possible choices for this task. This paper reports the results of the first exposure to a muon beam at CERN of a new hybrid chamber, obtained by coupling a GEM chamber and an emulsion detector. Thanks to the micrometric accuracy of the emulsion detector, the position resolution of the GEM chamber as a function of the particle inclination was evaluated in two configurations, with and without the magnetic field}

\keywords{Micropattern gaseous detectors; Neutrino detectors; Particle tracking detectors}

\begin{document}
\maketitle

\flushbottom

\section{Introduction}
\label{sec:intro}

SHiP (\textit{Search for Hidden Particles}) is a new experiment proposed at CERN in 2015 with the submission of a Technical Proposal \cite{TP} and an associated Physics Paper \cite{PP}.
The main goals of the experiment are the search for new physics at the intensity frontiers through the probe of different portals and the study of neutrino physics.

A 400 GeV/c proton beam produced at the CERN SPS will impinge on a hybrid target made of 58 cm of titanium-zirconium doped molybdenum (TZM) alloy and other 58 cm of tungsten, for a total of 10 interaction lengths. 
The length of the target is chosen in order to contain possible hadronic showers with minimum leakage. To completely absorb all the hadrons produced in proton interactions in the target, a block of 5~m of iron ($\sim$30 interaction lengths) is placed immediately downstream. 
A high muon flux, produced by the decays of the residual pions and kaons and by short-lived resonances, is expected. An active muon sweeper based on magnetic deflection of the muons in the horizontal plane is located immediately downstream of the hadron stopper in order to get rid as much as possible of background muons in the detector region.

A neutrino detector made of a 9.6 ton modular target is placed downstream of the muon shield and it is surrounded by a magnet providing a cylindrical magnetic field. The identification of muons and the measurement of their charge and momentum is provided by a muon spectrometer placed immediately downstream.
The modular structure of the neutrino target is made by alternating walls of target units, so-called bricks, and planes of electronic detectors.
Each brick is designed according to the Emulsion Cloud Chamber (ECC) technique \cite{ECC}, the same used by the OPERA \cite{OPERA} experiment for the discovery of tau neutrino appearance in a muon neutrino beam \cite{5tau}. 
A target brick is made of 57 emulsion films, each 300 $\mu$m thick, acting as tracking devices with a resolution of the order of the micrometer, interleaved with 56 lead plates with a thickness of 1~mm, serving as target material for neutrino interactions.
The micron accuracy of the emulsion films allows to detect both the $\nu_\tau$ interaction vertex and the $\tau$ lepton decay vertex.

To provide the time stamp of the neutrino interaction in the brick and link the muon track reconstructed in the emulsion with the muon spectrometer, each brick wall is followed by an electronic tracking chamber. Several options are currently under investigation in SHiP.
The Gas Electron Multiplier (GEM) \cite{GEM} is one of them. 

There are very strict requirements that these tracking chambers must satisfy.
They must cover the whole detector surface ($\approx$ 2 m$^2$) while keeping a constant distance of a few mm from the bricks so to ensure good planarity and uniform tracking performances.
Furthermore, the request to disentangle tracks from the same interaction coming from two separate vertices and the need of efficiently vetoing penetrating background muons requires these detector to have 100 $\mu$m position resolution on both coordinates and a high efficiency (>99$\%$) for angles up to 1 rad.

In preparation for the technological choice of electronic tracking chambers for the SHiP experiment, a new hybrid chamber obtained by coupling an electronic detector (GEM) and an emulsion detector was for the first time assembled and exposed to a muon beam at CERN. The results of this exposure are reported in this paper. With its micron position resolution, the emulsion detector has been also used to estimate the position resolution of the GEM chamber as a function of the track inclination in two running configurations, with and without magnetic field. 

\section{Experimental setup}
\label{sec:Setup}

The combined test of the detector composed of emulsion films and triple-GEM detector \cite{3GEM} was performed in October 2015 at the CERN SPS on the H4 beam line.

The required field of 1~T strength was supplied by the Goliath Magnet \cite{Goli}, located inside the PPE134 zone.
Its overall dimensions are 4.5$\times$3.6$\times$2.79 m$^3$ and it is composed of two horizontal parallel coils with a diameter of 2~m mounted in series and placed at a vertical distance of 1.05~m.
The produced magnetic field shows a cylindrical symmetry. The field is approximately constant in a 1~m diameter region, varying up to 30$\%$ when approaching the radius of the coils. It then drops rather fast, halving with respect to its maximum value at $\approx$ 1.2~m from the axis.

\begin{figure}[h]
\centering
\includegraphics[width=0.9\textwidth]{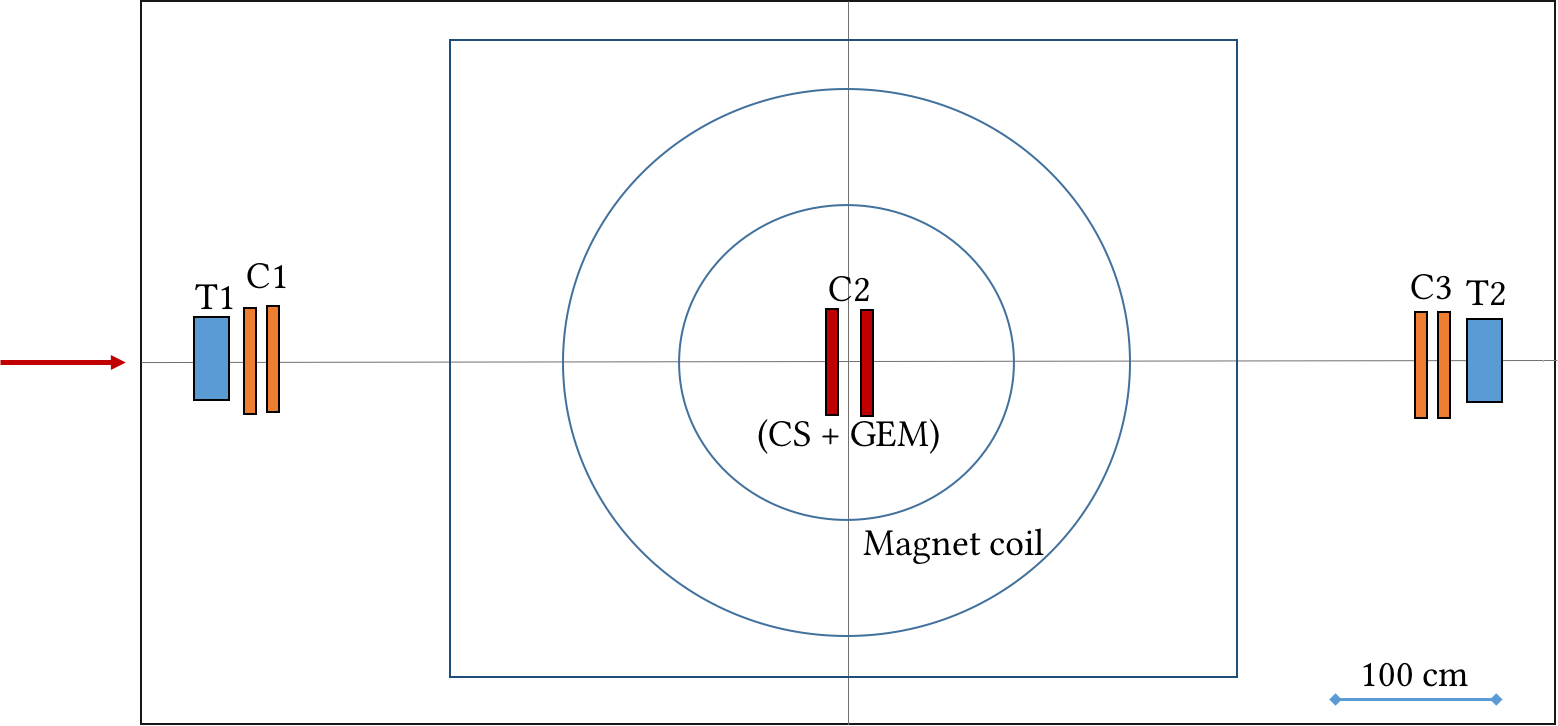}
\caption{Sketch of the experimental setup. T1 ant T2 are the scintillators used for trigger, C1 and C3 the pairs of triple GEM chambers used for tracking. C2 is the GEM-Emulsion system.}
\label{fig:Setup}
\end{figure}

\begin{figure}[h]
	\centering
	\includegraphics[width=0.5\textwidth]{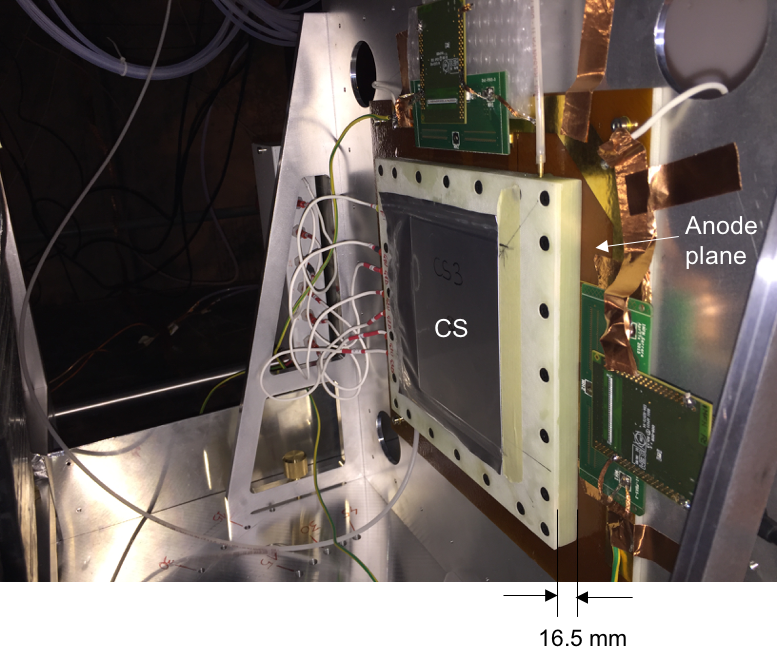} 
	\caption{Changeable Sheet (CS) envelope attached to the upstream side of the GEM detector.}
	\label{fig:CSsetup}
\end{figure}

The experimental setup is shown in fig.~\ref{fig:Setup} and is composed of:
\begin{itemize}
\item Scintillators with SiPM and PMT readout for the trigger; T1 upstream and T2 downstream. The trigger active area is about 4$\times$4 cm$^2$; 
\item Two tracking stations, C1 3.5 m upstream and C3 3.5 m downstream w.r.t. the test prototype; each station consists of two 20-cm-spaced XY triple-GEM chambers;
\item A triple GEM detector (C2) was placed inside the magnet cavity on a rotating platform to perform exposures at different angles. 
At 16.5~mm from the GEM anode, on its upstream surface, an emulsion doublet, that hereafter will be referred to as a Changeable Emulsion Sheet (called CS for historical reasons), was attached, as shown in fig.~\ref{fig:CSsetup}.
\end{itemize}

All GEM detectors (10$\times$10 cm$^2$) have been operated with an Ar/CO$_2$ 70:30 gas mixture and read-out by the SRS \cite{ReadOutGEM} featuring APV25 hybrid cards which provide an analog readout of the charge with 25 ns sampling time \cite{CMOSreadout}.

\begin{figure}[h]
	\centering
\subfloat[]{\includegraphics[width=0.47\textwidth]{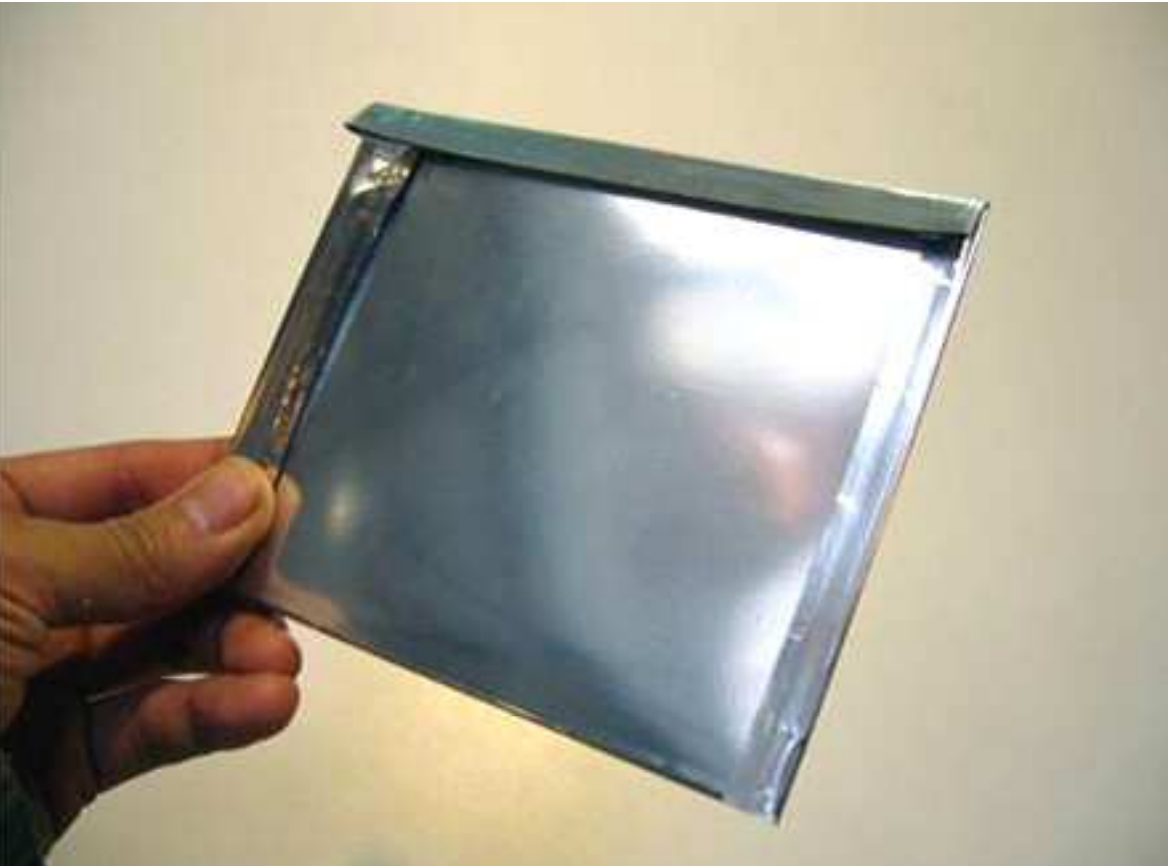} }\quad
\subfloat[]{\includegraphics[width=0.28\textwidth]{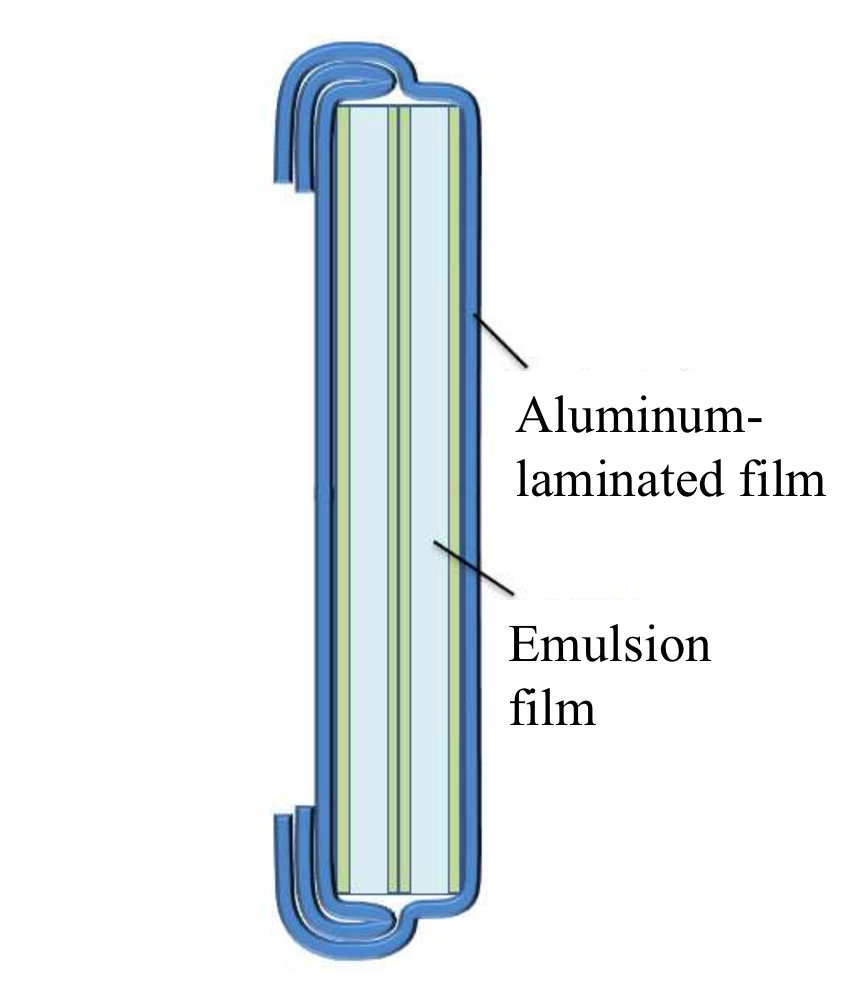} }
	\caption{Packed Changeable Emulsion Sheet (a) and schematic drawing of the cross-section (b).}
	\label{fig:CS}
\end{figure}

The CS is made of two adjacent emulsion films packed together using an aluminum-laminated film. This particular film grants the light tightness needed to prevent any damage to the emulsion films and it is approximately 100~$\mu$m thick. In fig.~\ref{fig:CS} a picture of a packed CS and a drawing of its cross-section is shown.

The three exposed CSs were assembled in the emulsion laboratory previously used for the OPERA experiment \cite{OPERA} at CERN which hosts a dark room equipped for the film handling. The envelope was packed under vacuum to prevent emulsion film displacement during the exposure for an overall thickness of the emulsion detector of 800$\pm$50 $\mu$m. Another advantage of this peculiar packaging is that it ensures a good rigidity under deformations. As shown in fig.~\ref{fig:CSsetup}, the centre of the CS was placed at 16.9 mm from the anode of the GEM detector.

Immediately after the exposure, the CSs were unpacked to reduce the accumulation of background
tracks from cosmic rays in the same geometry of the exposure. Indeed, cosmic ray tracks
accumulated in the time interval between production and CS assembly do not constitute background
because the films were arranged in a different configuration compared to the storage.

The films were later developed at Gran Sasso INFN Laboratories exploiting the emulsion facility used for the OPERA experiment.

The GEM-Emulsion system was exposed to 150 GeV/c muons produced at the SPS North Area.
Due to the high mean momentum of the muon beam, deflections due to the magnetic field as well as scattering effects are negligible in the region of the C2 system.
The beam intensity was measured using the coincidence of two scintillators placed immediately upstream of the magnet.

Given the micrometric accuracy of the emulsion detector, the resolution of the GEM-Emulsion system is fully dominated by the GEM:

\[\sigma_{GEM+CS} = \sqrt{\left(\sigma_{CS}\right)^2+\left(\sigma_{GEM}\right)^2} \sim \sigma_{GEM}\]

In the worst case of large angle and magnetic field, the GEM accuracy is expected to be always better than 400~$\mu$m.
Thereby, the exposure was planned with an integrated track density such that the average distance between adjacent tracks was $\sim 3 \sigma = 1.2$~mm. 
This sets the upper limit of 50 tracks/cm$^2$ to the track density for each angular region on the emulsion surface (10$\times$12.5 cm$^2$).
This average density of tracks still provides sufficient statistics to evaluate the parameters needed to define a common reference frame for the tracks in the CS and the hits in the GEM chamber.
To have less than 50 tracks/cm$^2$/angle with an exposure time of about ten minutes, the setup was displaced by about 10 cm with respect to the beam axis. 

Two sets of exposures were performed, one with the magnet off and the other one with the field strength set at 1 T, as expected in the SHiP neutrino detector. In each exposure the GEM-Emulsion system was rotated in oder to have 0$^\circ$, 7.5$^\circ$, 15$^\circ$ and 30$^\circ$ incident angles from the C2 normal. In the configuration with the magnet on, an additional exposure at 45$^\circ$ was performed.
This allowed to test the matching between the emulsion and the GEM detector in different experimental conditions.

\section{The triple-GEM detector}
\label{sec:GEM}

\begin{figure}
\begin{center}
\begin{minipage}[c]{.45\textwidth}
\centering
\includegraphics[width=1\textwidth]{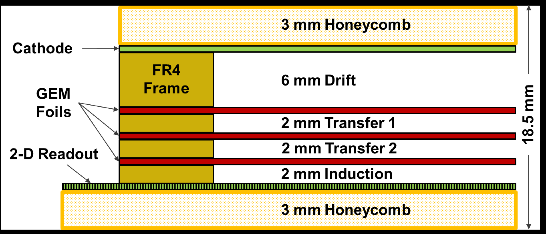}
\caption{Cross section of a triple-GEM detector.}
\label{fig:GEMxs}
\end{minipage}%
\hspace{5mm}%
\begin{minipage}[c]{.45\textwidth}
\centering
\includegraphics[width=.60\textwidth]{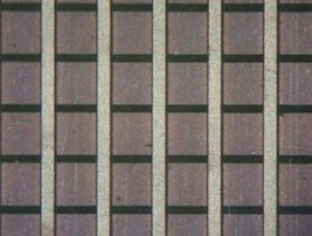}
\caption{Detail of the readout circuit. The X and Y strips have the same pitch but different widths to equally share the signal since the strips layers are at different distance from the last GEM foil.}
\label{fig:ReadOut}
\end{minipage}
\end{center}
\end{figure}

The Gas Electron Multiplier (GEM) \cite{GEM} is made of a 50 $\mu$m thick polyimide foil, copper clad on each side and chemically etched with a high density of holes (70 $\mu$m diameter, 140 $\mu$m pitch). When a voltage difference of typically 400-500 V is applied between the two copper faces, an electric field as high as 100 kV/cm is created within the holes,  that act as multiplication channels for the ionization electrons produced in the drift/conversion gap of the detector. A triple stages multiplication structure allows to reach a gain as high as 10$^4$ while minimizing the discharge probability \cite{3GEM}. Figure \ref{fig:GEMxs} shows a cross-section of the triple-GEM detector used in the test. The large drift gas gap of 6 mm, providing a high primary ionization, allows a full detector efficiency  as well as an easy reconstruction  of the track angle when the detector is operated in micro-TPC mode \cite{muTPC1, muTPC2, muTPC3}. The readout plane is a multi-layer circuit on a polyimide substrate, patterned with an XY structure of copper strips engraved at two different levels (fig. \ref{fig:ReadOut}). A strip pitch of 650 $\mu$m has been adopted for all GEM chambers.

\subsection{Data reconstruction}

For particles crossing the triple-GEM detectors, the position of the cluster of the fired strips is reconstructed by the weighted average of the charge. This method, called charge centroid (CC), allows to improve the spatial resolution with respect to the digital readout, where the spatial resolution is simply given by the strip multiplicity divided by $\sqrt{12}$.

For particles crossing the detector orthogonally and without magnetic field the charge distribution depends mainly on the electron diffusion within the gaps, that is driven by the gas mixture and the electric fields inside the chamber. 

\begin{figure}[h]
	\centering
	\includegraphics[width=0.5\textwidth]{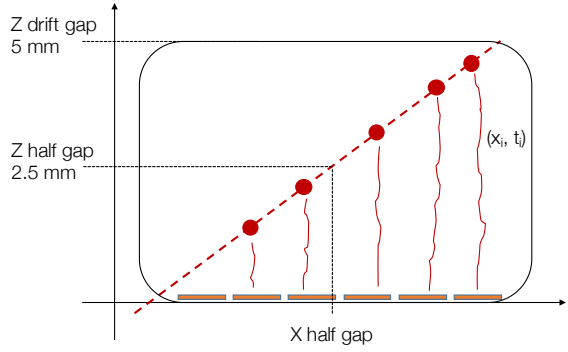} 
	\caption{Sketch of a non-orthogonal track inside the drift/conversion gap, showing the mechanism that increases the strip multiplicity of strip clusters.}
	\label{fig:GEMtrk}
\end{figure}

\begin{figure}[h]
	\centering
	\includegraphics[width=0.8\textwidth]{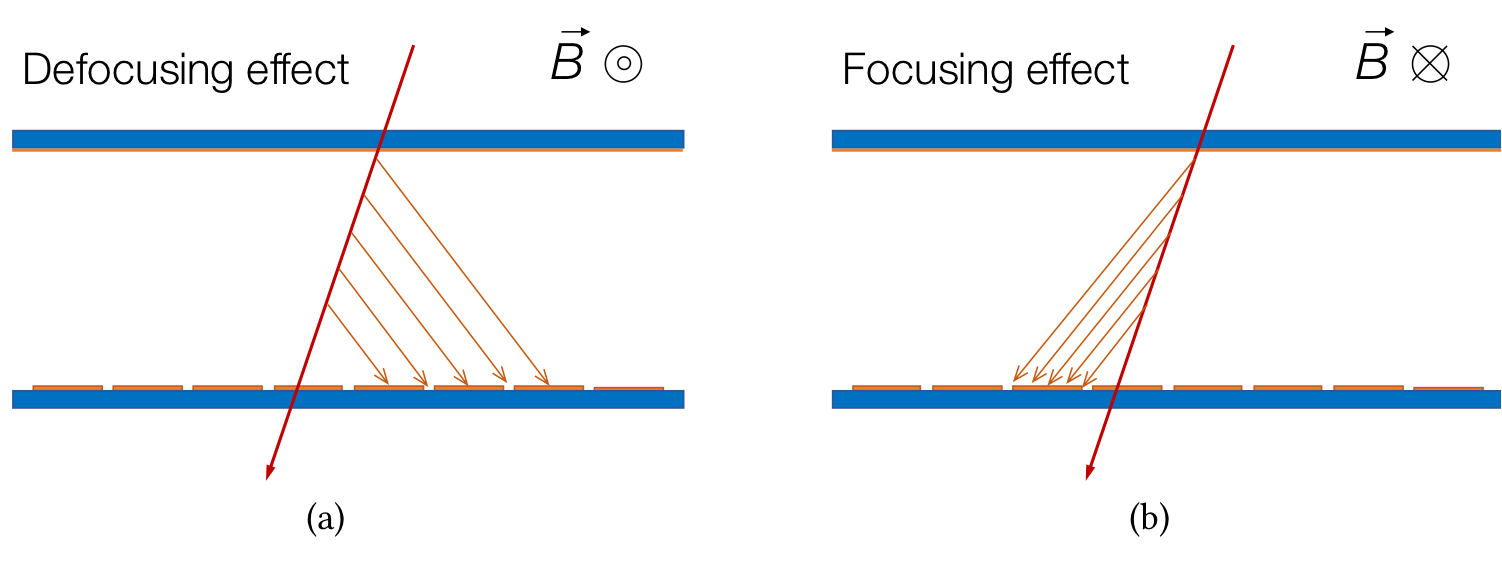} 
	\caption{Sketch of defocusing (a) and focusing (b) configuration on a typical MPGD.}
	\label{fig:Focus}
\end{figure}

When the incident angle of the particle is not orthogonal to the detector, the ionization electrons drifting along the direction of the electric field lines, increase the strip multiplicity of clusters (see fig.~\ref{fig:GEMtrk}) by spreading the charge. 
When a strong magnetic field is applied orthogonally to the drift field, the Lorentz force displaces the electrons: the effect is a heavy smearing and deformation of the electron charge shape at the anode. The result is an overall deterioration of the spatial resolution (fig.~\ref{fig:Focus} a).

A peculiar configuration occurs when the angle of the incident particle is near the Lorentz angle, which depends on the gas mixture and the electric fields (in our case about 15$^\circ$).

From the point of view of the track reconstruction, this configuration is equivalent to a perpendicular track in absence of magnetic field, where the clusters have minimal spread. This special configuration is generally referred as focusing configuration (fig. \ref{fig:Focus} b).

\section{Emulsion: detector and data analysis}
\label{sec:Emu}
Nuclear emulsion films are made of \emph{AgBr} crystals dispersed in a gelatine binder. \emph{AgBr} crystals are semiconductors with a band-gap of 2.6~eV. The transit of a charged particle creates electron-hole pairs producing \emph{Ag} metal atoms acting as latent image centres. Through a process called development, the number of silver atoms multiplies and the size of silver grains increases (from 0.2~$\mu$m to 0.6~$\mu$m) becoming visible with an optical microscope.

\begin{figure}[h]
	\centering
	\includegraphics[width=0.5\textwidth]{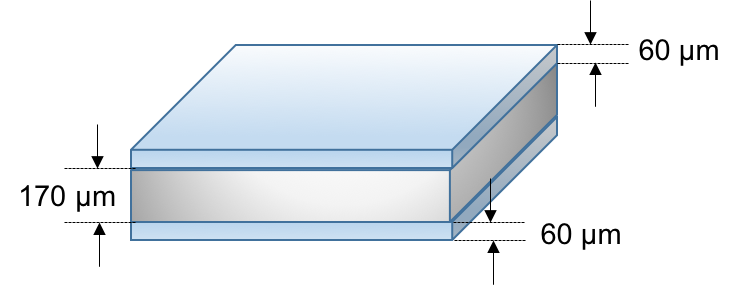} 
	\caption{Sketch of the transversal section of an emulsion film.}
	\label{fig:emulsion}
\end{figure}

Each emulsion film is made of two, 60~$\mu$m thick, active layers poured on both sides of a 170~$\mu$m thick polystyrene base (see fig.~\ref{fig:emulsion}), for a total thickness of 290~$\mu$m, while the transverse sizes is 10$\times$12.5~cm$^2$. 
The films used for this test beam have been produced at Nagoya University with a new emulsion gel different from the one used in the OPERA experiment with a higher number of grains (50 grains/100 $\mu$m vs. 30 grains/100 $\mu$m produced by a minimum ionising particle).

The emulsion films have been analysed using an upgrade \cite{ESSupgrade} of the OPERA European Scanning System (\emph{ESS}) \cite{ESS}, made by the Naples emulsion scanning group.
The ESS is a Cartesian robot with a digital camera for image grabbing mounted on the optical axis (Z) of the microscope and connected to a vision processor. A vacuum pumping system holds in place the films which are placed on a motor driven horizontal stage able to move along both the x and y axes with a maximum excursion of 20.5 cm. Details are reported in \cite{ESS} and in \cite{ESSupgrade}.

The use of a faster camera with smaller sensor pixels and a higher number of pixels combined with a lower magnification objective lens, together with a new software LASSO \cite{lasso} has allowed to increase the scanning speed to 84 cm$^2$/hour, more than a factor of four larger than before.

The lens of the microscope guarantees a submicron resolution and, having a working distance in Z longer than 300 $\mu$m, enables the scanning of the lower part of the emulsion film without needing to tip it over. 
To make the optical path homogeneous in the film and the microscope optics, an immersion lens in an oil with the same refraction index of the emulsion is used. 
A single field of view is 800$\times$600 $\mu$m$^2$; larger areas are scanned by repeating the data acquisition on a grid of adjacent fields of view.

The images grabbed by the digital camera with a stop-and-go algorithm are sent to a vision processing board in the control workstation to suppress noise.
Three-dimensional sequences of aligned clusters (grains) are recognised and reconstructed by the CPUs of the host workstation. 
These sequences identify what is called a micro-track in the emulsion layer if formed by at least 6 or at the most 16 clusters.
The position assigned to a micro-track is its intercept with the nearest plastic base surface.

An area of 8$\times$8 cm$^2$ has been scanned on each emulsion layer of each film.  To avoid fiducial volume effects, 1~cm was left from the border of the GEM chamber.

The track recognition procedure is a quite complex process executed by the LASSO software tracking module. The steps of the algorithm for the micro-track reconstruction are described in \cite{lasso}.
Once micro-tracks have been reconstructed, the following steps of the analysis have been performed with a dedicated offline software \cite{fedra}. 
First a procedure aiming at the correction of the reduction of the thickness of the emulsion layer  (\textit{shrinkage}), arising after development due to the dissolution of silver halide in the fixing phase, was applied. Subsequently, the two corresponding track segments in the emulsion layers on both sides are linked forming a so-called base track (BT): this step is called \emph{linking}. 
This is important to reduce the instrumental background due to fake combinatorial alignments and to increase the precision on the reconstruction of the track angle, minimising distortion effects.

The full-volume wide reconstruction of particle tracks requires connecting base-tracks in the two consecutive films of the CS doublet. 
In order to define a global reference system, a set of affine transformations has to be computed to account for the different reference frames used for data taken in different plates. Affine transformations correct also misalignments.
An affine transformation is given by:

\begin{equation}
 \begin{pmatrix} 
 x_M  \\ y_M 
 \end{pmatrix}
 =
\begin{pmatrix} 
a_{11} && a_{12} \\ a_{21} && a_{22} 
 \end{pmatrix}
\begin{pmatrix} 
x_t  \\ y_t
\end{pmatrix}
 + 
\begin{pmatrix} 
x_0 \\ y_0
\end{pmatrix}
\label{eq:affinetransf}
\end{equation}

where x(y)$_M$ are the measured coordinates of a point, while x(y)$_t$ are the expected ones. The z coordinate is not affected by these misalignments and, consequently, the transformation matrix is a 2$\times$2 matrix and the total number of parameters to be found is six: the four elements of the matrix and x$_0$ and y$_0$, the offset parameters, responsible for the translations.

\begin{figure}[h]
\centering
	\includegraphics[width=0.45\textwidth]{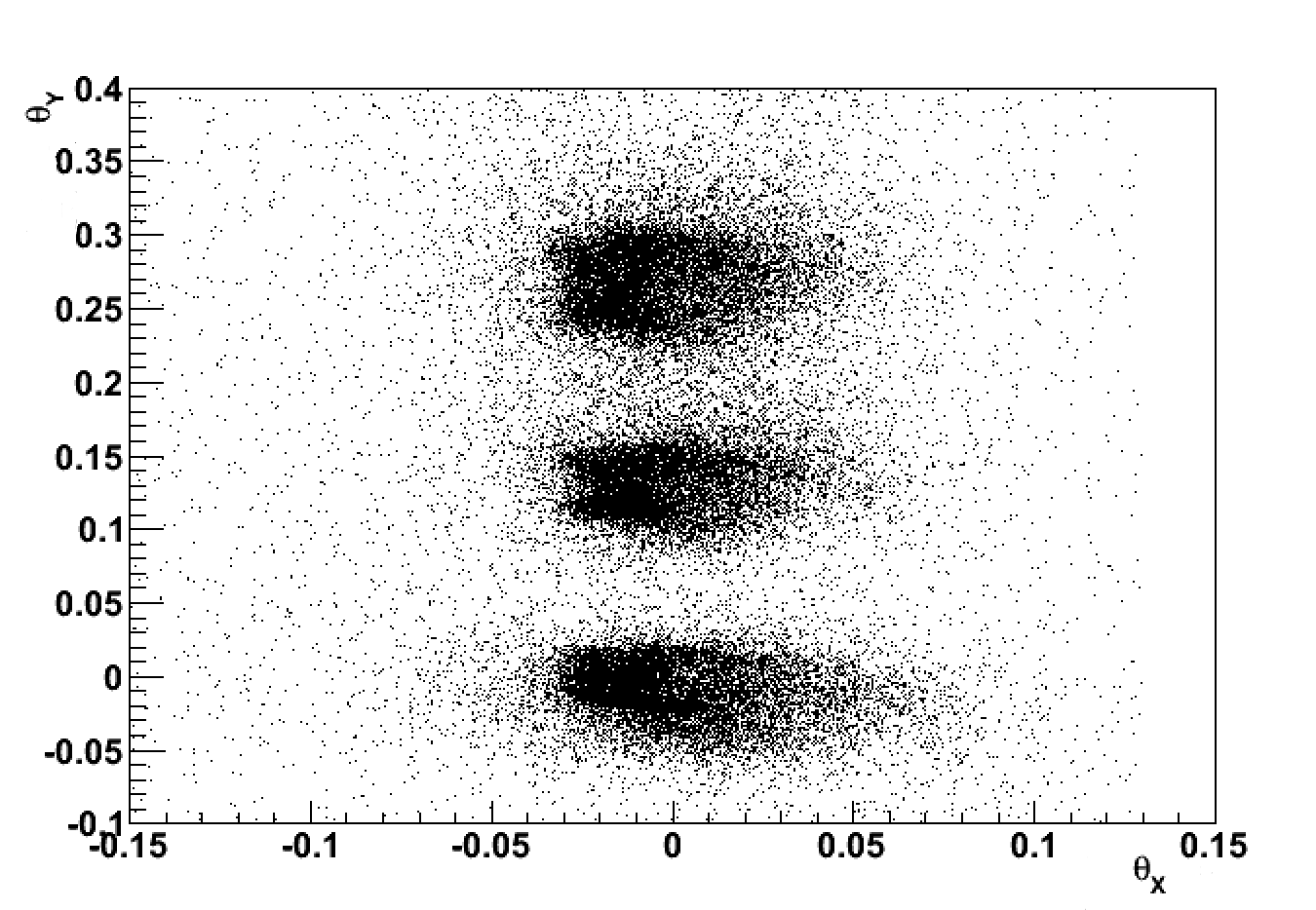} 
	\caption{Y slope ($\tan \theta_Y$) vs. X slope ($\tan \theta_X$) of the reconstructed tracks in the CS. The three darker regions identify the angles at which the exposure was performed within the errors in positioning.}
	\label{fig:distrib}
\end{figure}

After the alignment phase, tracks in the CS doublet have been reconstructed. In fig.~\ref{fig:distrib} the Y slope ($\tan \theta_Y$) versus the X slope ($\tan \theta_X$) of the reconstructed tracks in one of the exposed CS is reported. It can be clearly seen that the reconstructed tracks of the beam identify three different regions in the $\theta_Y - \theta_X$ plane corresponding to the different angles at which the exposure was performed. On the other hand, the background tracks are uniformly distributed in the $\theta_Y - \theta_X$ plane. These background tracks are the cosmic rays accumulated throughout the whole exposure, from the assembling up to the dismantling of the CS envelope. The central value of the three darker regions in fig.~\ref{fig:distrib} is compatible with the nominal one within errors in the positioning.

\section{Pattern matching results}
\label{sec:Res}

Unlike electronic detectors, emulsion based detectors are time insensitive, recording all charged tracks which cross them. Therefore, it is not possible to match tracks on the CS doublet and on the GEM chamber by using the timing at which the tracks have produced the corresponding hits. 

The only way to couple the tracks reconstructed in the two detectors is to use the information on the position of the hits. The ensemble of the $x-y$ positions of the tracks on each detector is called pattern.

The pattern matching is performed in two different steps: a rough alignment and a finer one. 

\begin{figure}[h]
	\centering
\includegraphics[width=0.45\textwidth]{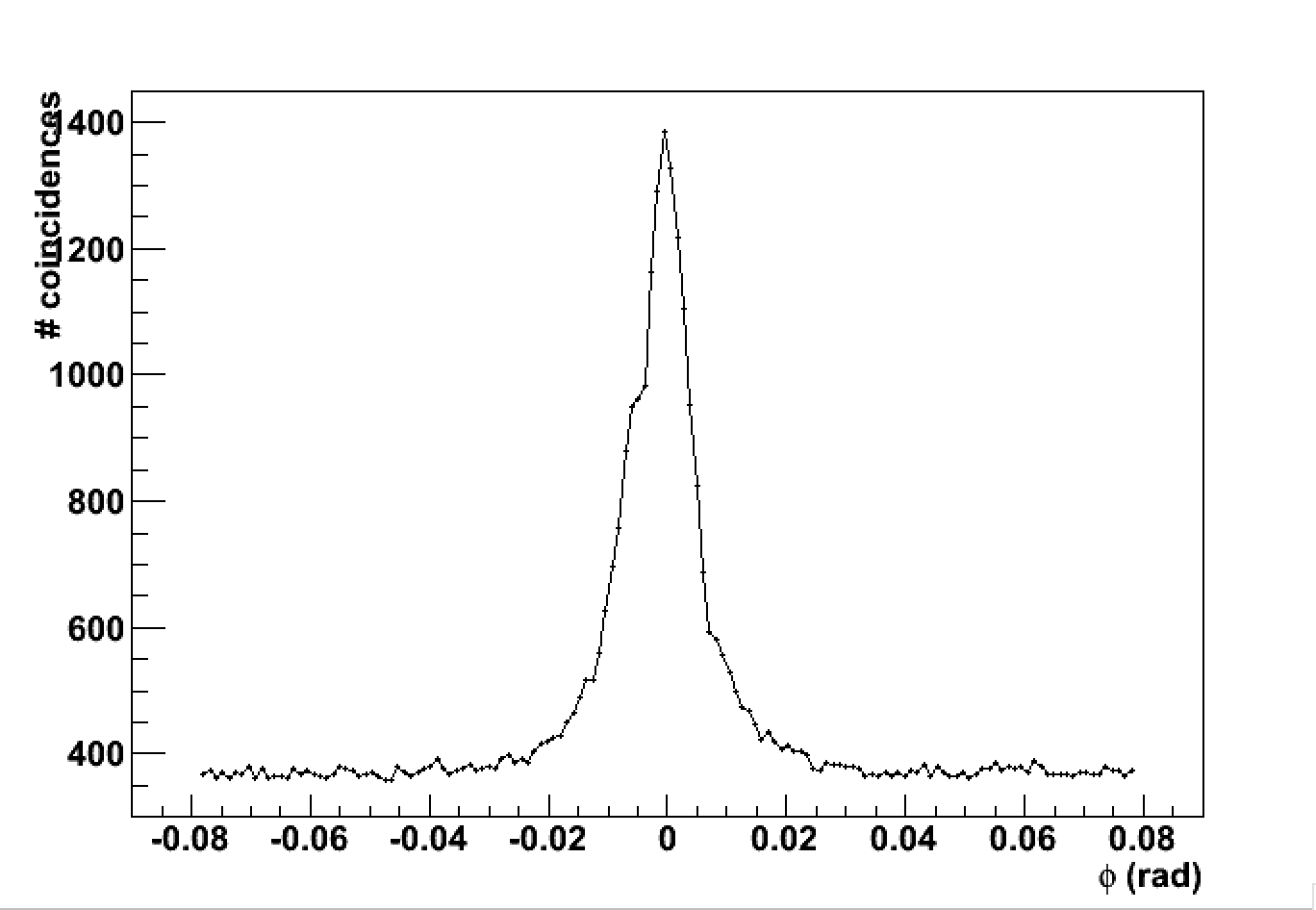} 
	\caption{Number of coincidences computed at different $\phi$ rotation angles between the CS tracks and the GEM hits. The distribution is peaked at the $\phi$ value to be used as a start for the alignment procedure between CS and GEM.}	
	\label{fig:align}
\end{figure}

The rough alignment aims at identifying both the rotation angle around the z axis and the two offsets in position ($\Delta x$, $\Delta y$) which define the level of displacement between the pattern of tracks on the CS doublet and that on the GEM detector. 
To this aim, the two patterns are rotated around their center of gravity with an angular step $\Delta \phi \sim$ 1~mrad. For each rotation angle, all the possible $\Delta x$, $\Delta y$ (up to 5 mm) are considered with steps of 500~$\mu$m (equivalent to the maximum expected resolution of the GEM-Emulsion system).
At each step, the distance between all the tracks belonging to the two patterns is evaluated. If the distance between two tracks is below 500~$\mu$m, they are said to be coincident. The number of coincidences between the two patterns is thus counted for each offset. 
The maximum number of coincidences defines the best $\Delta x$, $\Delta y$ combination for the considered rotation angle. 
When plotting the maximum number of coincidences found at each step as a function of the rotation angle, $\phi$, the distribution peaks at a certain $\phi$ value, see fig.~\ref{fig:align}. This is the rotation angle to be used as a start for the finer alignment procedure between CS and GEM.

\begin{figure}[h]
	\centering
\includegraphics[width=0.45\textwidth]{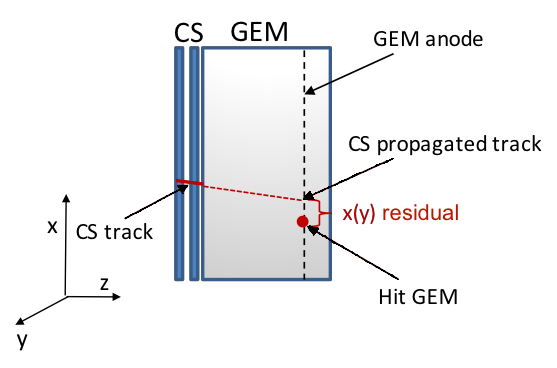} 
	\caption{Scheme of the GEM-Emulsion system setup. The axes of the reference frame are taken in such a way that the z axis is parallel to the beam, the x axis is parallel to the direction of the magnetic field.}
	\label{fig:hitdisplacement}
\end{figure}

The finer alignment uses affine transformations, as those described in paragraph \ref{sec:Emu}.
By applying to one of the two track patterns the affine transformation evaluated as described above, a set of track segments in both the emulsion and in the GEM sub-detectors is defined in a common reference system. Those track segments have been used to estimate the tracking accuracy of the whole detector. The track segment reconstructed in the emulsion has been extrapolated to the nominal position of the GEM anode and its position has been compared with the GEM hit as shown in fig. \ref{fig:hitdisplacement}. The Gaussian fit of the residuals provides the position accuracy.

\begin{figure}[h]
 \centering	
 \includegraphics[width=0.55\textwidth]{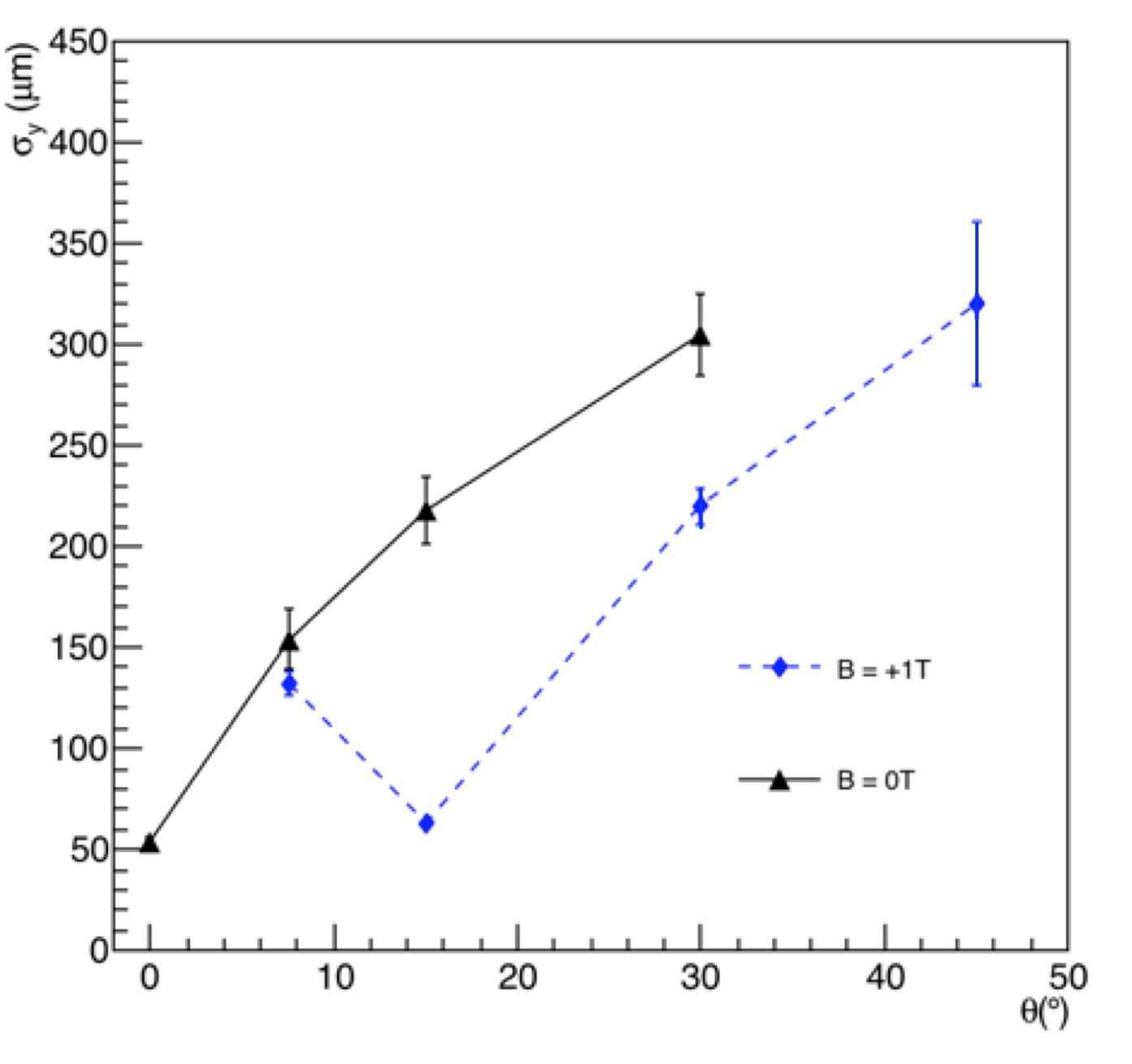}
 \caption{Resolution of the GEM as a function of the incident angle and for different values of the magnetic field.}
 \label{fig:summary}
 \end{figure}

\begin{figure}[h]
	\centering
\subfloat[]{\includegraphics[width=0.45\textwidth]{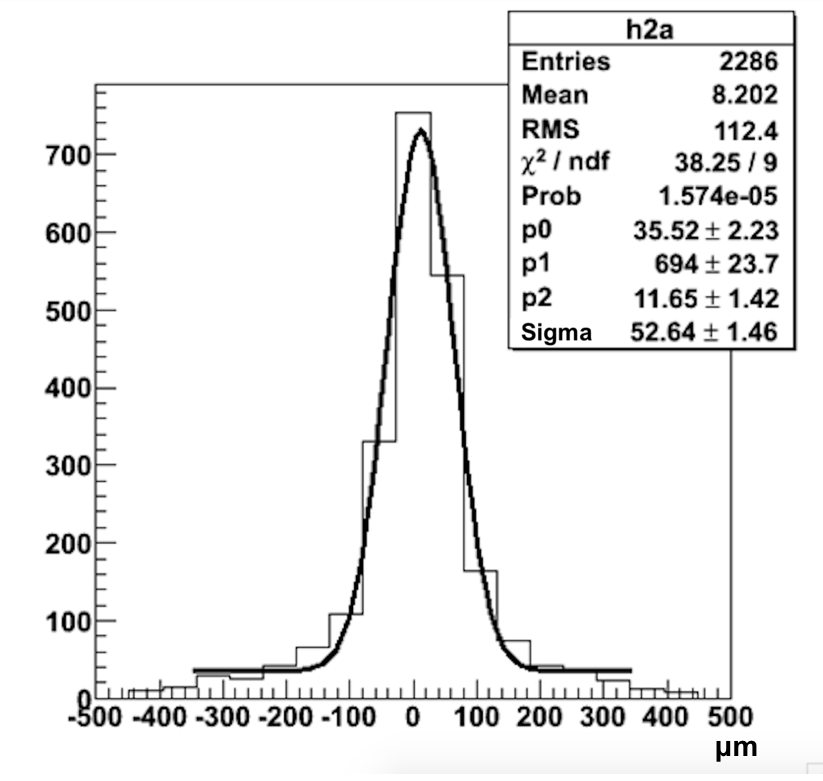}}\quad
\subfloat[]{\includegraphics[width=0.45\textwidth]{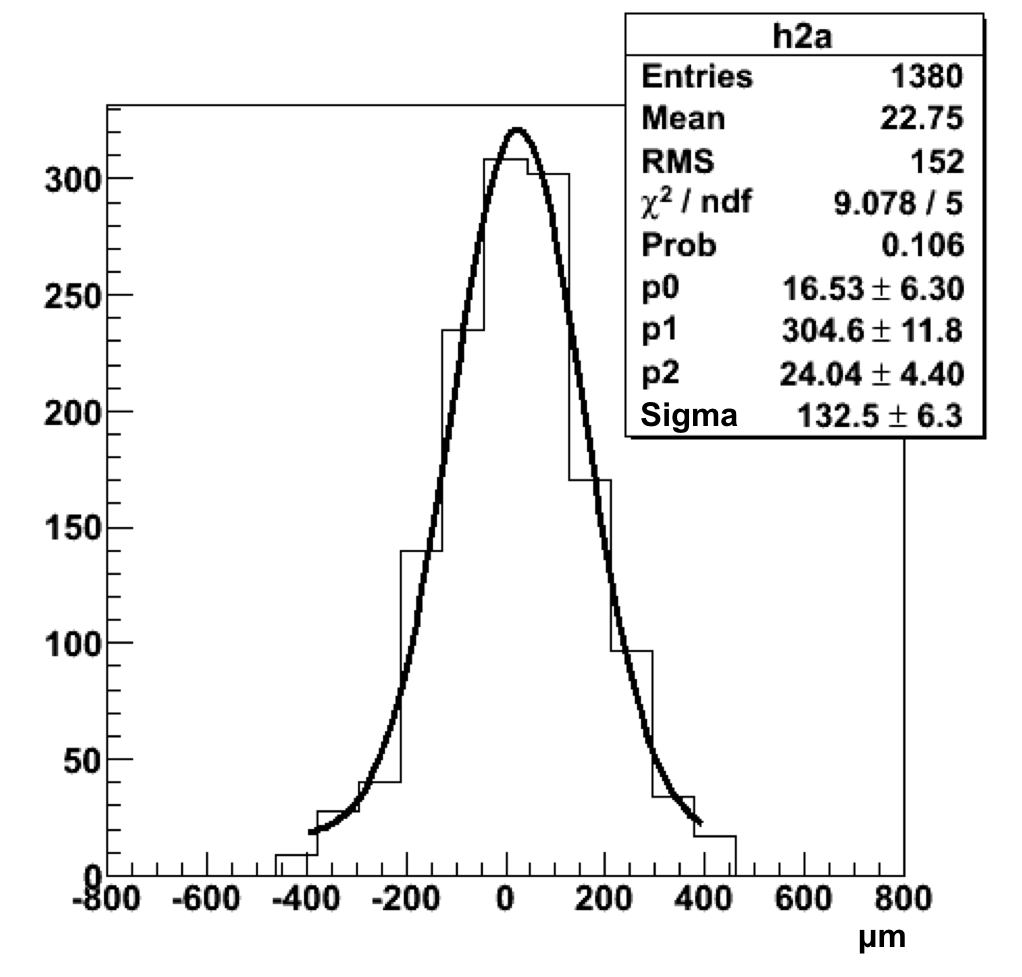} }\\
\subfloat[]{\includegraphics[width=0.45\textwidth]{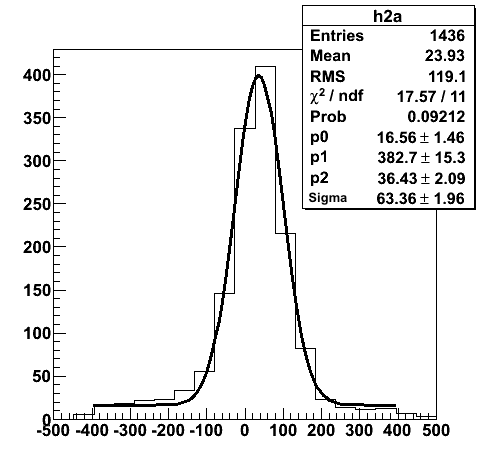} }
	\caption{Residuals between y coordinate of GEM hits and CS tracks for $\theta = 0^{\circ}$ and B = 0 (a), $\theta = 7.5^{\circ}$ and $\theta = 15^{\circ}$ and B = 1 T (b,c).}
	\label{fig:phiDistrib}
\end{figure}

In the configuration with $B = 0$ and an incident angle of the particle $\theta$ = 0$^{\circ}$, the GEM detector has a spatial resolution of $\sigma_{B=0, \theta=0}$ = (54$\pm$2) $\mu$m. The resolutions of GEM for configurations with different particle angles and values of the magnetic field are reported in table \ref{tab:results} and in fig.\ref{fig:summary}. 
In fig.~\ref{fig:phiDistrib} the Gaussian widths of fitted residuals distributions for $\theta = 0^{\circ}$ and $B = 0$, $\theta = 7.5^{\circ}$ and $\theta = 15^{\circ}$ both with $B = 1$~T are reported.

\begin{table}[h]
	\begin{center}
		\begin{tabular}{ccc}
			\toprule
								&   $\sigma (\mu m)$ &  \\
			\toprule
			& B=0 & B=+1T\\
			\midrule
			$\theta$ = 0$^{\circ}$ & 54$\pm$2 &\\
			\hline
			$\theta$ = 7.5$^{\circ}$ & 154$\pm$15 &  132 $\pm$ 6 \\
			\hline
			$\theta$ = 15$^{\circ}$ & 218$\pm$17&   63$\pm$2\\
			\hline
			$\theta$ = 30$^{\circ}$ & 305$\pm$20 &  $220\pm9$\\
			\hline
			$\theta$ = 45$^{\circ}$ &  & $320\pm40$\\
			\bottomrule
		\end{tabular}
	\end{center}
	\caption{Measured GEM position resolution for different incidence angles of particle from the CS normal and strengths of the magnetic field.}
	\label{tab:results}
\end{table}

As it can be inferred from the reported results, the resolution performances of the GEM detector depend strongly both on the angle of the track and on the magnetic field.
A non-zero incident angle of the incoming track produces a broadening of the charge distribution on the readout strips, thus worsening the estimate of the position. 

The magnetic field strength of 1 T produced a Lorentz angle that counters the effect of an increasing incident angle of the track. 
This is visible already for an inclination of 7.5$^\circ$ where $\sigma$ = $(132\pm6)$ $\mu$m and it is even more evident at 15$^\circ$ where the resolution is $\sigma$ = $(63\pm2)$ $\mu$m, compatible with the intrinsic resolution in absence of magnetic field.
Further increasing the particle incident angle leads to a broader charge distribution on the readout strips and thus to a degradation on the position resolution up to $(320\pm40)$ $\mu$m for $\theta = 45^\circ$. 
Therefore 15$^\circ$ corresponds to a rather complete compensation of the Lorentz angle.
We note how, within the errors, the two curves in fig.~\ref{fig:summary} for $B=1$~T and $B~=~0$ are almost parallel for $\theta > 15^{\circ}$, with the two configurations shifted by a 15$^\circ$ phase corresponding to the Lorentz angle.

 
\section{Conclusions}
\label{sec:end}

For the first time a hybrid chamber made of a GEM chamber and an emulsion detector was assembled
and tested using a 150 GeV/c muon beam at the SPS accelerator at CERN. The hybrid chamber was
obtained by attaching a doublet of emulsion films (CS) to the upstream surface of a GEM detector.
This test is required for the technological choice of an electronic detector to be used in the neutrino
detector of the SHiP experiment. It proved the capability to reconstruct the trajectory of particles
crossing this hybrid chamber with an accuracy better than 100 $\mu$m.

Different exposures were performed at different incident angles of the beam on the detector detector, with and without magnetic field. The field of the magnet was set to 1~T as expected in the SHiP neutrino detector. 
The trajectories of several thousands of particles have been reconstructed in the hybrid chamber; the corresponding position resolution, dominated by the space resolution of the GEM chamber, has been studied as a function of the track inclination and of the magnetic field.

The measured GEM position resolution ranges from $\sigma$ = (54$\pm$2) $\mu$m at zero angle in absence of field up to $(320\pm40)$ $\mu$m for $\theta = 45^\circ$ and magnetic field strength of 1 T.
As expected, the best result is obtained without magnetic field and with a zero incident angle of the beam on the detector.  The position resolution degrades by a factor 2 every $\approx$ 10$^{\circ}$, but a compensation due to the drifting of electrons at the Lorentz angle has been found at $\theta = 15^\circ$ when $B = 1$~T.

The resolution at $\theta = 0^{\circ}$ and without magnetic field complies with the needs of the SHiP experiment.
The degradation of the resolution for inclined tracks spoils significantly the performances of the GEM detector. 
Nevertheless, the micro-TPC mode is being developed for the reconstruction of tracks in the GEM detector. The micro-TPC mode could indeed considerably improve the position resolution also for inclined tracks and in presence of magnetic field. This new option will thus be tested in the future.

\section{Acknowledgments}
This project has received funding from the European Union's Horizon 2020 research and innovation program under grant agreement no 654168.

\end{document}